\documentstyle[preprint,aps]{revtex}
\begin{document}
\draft
\title{A variational perturbation scheme for many-particle systems in the 
       functional integral approach}
\author{Sang Koo You$^1$, Chul Koo Kim$^1$, Kyun Nahm$^2$ and 
        Hyun Sik Noh$^3$}
\address{1. Institute of Physics and Applied Physics, Yonsei University, Seoul
            120-749, and Center for Strongly
            Correlated Materials Research, Seoul National University, Seoul
            151-742, Korea}
\address{2. Department of Physics, Yonsei University, Wonju 220-710, Korea}
\address{3. Hanlyo University, Kwang-Yang 545-800, Korea}
\maketitle

\begin{abstract}
A variational perturbation theory based on the functional integral approach
is formulated for many-particle systems. Using the variational action obtained
through Jensen-Peierls' inequality, a perturbative expansion scheme for the
thermodynamic potential is established. A modified Wick's theorem is obtained 
for the variational perturbation expansions. This theorem allows one to carry out
systematic calculations of higher order terms without worrying about the double
counting problem. A model numerical calculation was carried out on 
a nucleon gas system interacting through the Yukawa-type 
potential to test the efficiency of the present method.  
\end{abstract}


\section{Introduction}

The perturbative approach has been proven widely successful when the interactions 
between the particles were weak[1-3]. However, when the correlations between the
particles are strong, the perturbation approaches are less than satisfactory.
In order to overcome such difficulties, many nonperturbative schemes 
including  variational calculation have been
developed [4-6]. 
Although the variational procedure provides a simple and convenient
 scheme to calculate the 
physical quantities, it has a serious drawback. Since, it is a basically
one-shot process, systematic improvement over the obtained results is not
easy to implement.

In this paper, we present a functional integral formulation of variational
perturbation scheme for many-particle systems. 
In fact, this concept has already been used in connection with various research 
interests such as lattice dynamics, relativistic field theories,
and quantum mechanics [7-12]. However, to the authors' knowledge, it has
never been applied to nonrelativistic many-particle systems.
 For this purpose,
 we first derive the vatiation principle for the partition function and
the thermodynamic potential in the functional integral formalism
using Jensen-Peierl's
inequality [13,14]. Then, this principle is used to obtain a variational action
for the functional integral. The difference between the true and the
variational action provides a renormalized perturbation. Application of the
Wick's theorem on this renormalized perturbation leads into cancellation
of intracell joining diagrams. It will be shown that this modified Wick's
theorem allows us to carry out systematic improvements on the variational
result without worrying about the problem of double counting.
As a test on the efficiency of the present scheme, we calculate up to the second
order contributions to the ground state energy of interacting nucleon gas 
through the Yukawa-type potential using both
the conventional and the present variational
perturbation scheme.

\section{The Variational Principle in the Functional Integral Approach}

The variational principle for the partition function and the thermodynamic
potential can be obtained using Jensen-Peierls' inequality [13,14], which is
given as follows for Hermitian operators $A_0$ and $A_1$ ;
\begin{eqnarray}
  \frac{{\rm Tr} e^{-(A_0 + A_1)}}{{\rm Tr} e^{- A_0}} \geq
   e^{- \langle A_1 \rangle_0} \ ,
\end{eqnarray}
where
\begin{eqnarray}
  \langle A_1 \rangle_0 =
  \frac{{\rm Tr} A_1 e^{- A_0 }}{{\rm Tr} e^{- A_0}} \ . \nonumber
\end{eqnarray}
When $A_0$ and $A_1$ are normal-ordered operators composed of creation 
and annihilation operators, $A_0 = \ :A_0 ( \{ a_{\alpha}^{\dagger},a_{\alpha} \} ) :$ 
and $A_1 = \ : A_1 ( \{ a_{\alpha}^{\dagger},a_{\alpha} \} ) :$ 
, the trace operation is expressed
in the coherent state functional integral form as follows [3],
\begin{eqnarray}
{\rm Tr} e^{-A_0 } 
= \int D [\bar{\psi} \psi] e^{-\int d\tau \left[\sum_{\alpha} \bar{\psi}_{\alpha}(\tau)
\partial_{\tau} \psi_{\alpha}(\tau) + A_{0} \left( \{ \bar{\psi}_{\alpha}(\tau),
\psi_{\alpha}(\tau) \} \right) \right] } \ .
\end{eqnarray}
Defining the generating function $K[\bar{\eta},\eta ; A_0 ]$ with source variables $\bar{\eta}$ and $\eta$,
\begin{eqnarray}
  K[\bar{\eta},\eta ; A_0 ] 
 \equiv \int D[\bar{\psi} \psi] e^{-\int d\tau \left[\sum_{\alpha} \bar{\psi}_{\alpha}(\tau)
\partial_{\tau} \psi_{\alpha}(\tau) + A_{0} \left( \{ \bar{\psi}_{\alpha}(\tau),\psi_{\alpha}(\tau) \} \right)
+ \sum_{\alpha} \psi_{\alpha}(\tau) \bar{\eta}_{\alpha}(\tau)
+ \sum_{\alpha} \eta_{\alpha}(\tau)\bar{\psi}_{\alpha}(\tau)  \right]} \ ,
\end{eqnarray}
we can express the above traces as follows,
 \begin{eqnarray}
&&{\rm Tr} e^{-A_0 } = K[\bar{\eta},\eta ; A_0 ] |_{\bar{\eta}, \eta =0} \ ,  \\
&&{\rm Tr} e^{-(A_0  + A_1)} = e^{-\int d\tau A_{1} 
\left( \left\{ \zeta \frac{\partial}{\partial
\eta_{\alpha}(\tau)}
,\frac{\partial}{\partial \bar{\eta}_{\alpha}(\tau) } \right\} \right)}
 K[\bar{\eta},\eta ; A_0 ] |_{\bar{\eta}, \eta =0} \ , \\
&&{\rm Tr} A_1   e^{-A_0 
} = \int d\tau A_{1} \left( \left\{ \zeta \frac{\partial}{\partial
\eta_{\alpha}(\tau)}
,\frac{\partial}{\partial \bar{\eta}_{\alpha}(\tau) } \right\} \right)
K[\bar{\eta},\eta ; A_0 ] |_{\bar{\eta}, \eta =0} \ ,
\end{eqnarray}
where $\zeta $ is $-1 (+1)$  and $\eta$ 
and $\bar{\eta}$ are Grassmann (complex) variables for fermion (boson) field.
Using the shorthand notation,
\begin{eqnarray}
\left \langle F \left( \left\{ \frac{\partial}{\partial \eta_{\alpha}(\tau)}
,\frac{\partial}{\partial \bar{\eta}_{\alpha}(\tau)} \right\} \right) \right \rangle_{A_0 } \equiv
\frac{ F\left( \left\{ \frac{\partial}{\partial \eta_{\alpha}(\tau)}
,\frac{\partial}{\partial \bar{\eta}_{\alpha}(\tau)} \right\} \right) K[\bar{\eta},\eta ; A_0 ]_{\bar{\eta}, \eta =0}}
{K[\bar{\eta},\eta ; A_0 ]_{\bar{\eta}, \eta =0}} \ \ ,
\end{eqnarray}
we can express Jensen-Peierls' inequality in the functional integral expression,
\begin{eqnarray}
\left \langle   e^{-\int d\tau A_{1} \left( \left\{ \zeta \frac{\partial}{\partial
\eta_{\alpha}(\tau)}
,\frac{\partial}{\partial \bar{\eta}_{\alpha}(\tau) } \right\} \right)}   \right \rangle_{A_0 }  \geq
e^{-\left \langle \int d\tau A_{1} \left( \left\{ \zeta \frac{\partial}{\partial
\eta_{\alpha}(\tau)}
,\frac{\partial}{\partial \bar{\eta}_{\alpha}(\tau) } \right\} \right)   \right \rangle_{A_0 } } \ .
\end{eqnarray}
In order to apply this inequality to the partition function, we rewrite the action
$S$ in the form $S = S_0 (\lambda) + (S - S_0 (\lambda))$, where $S_0 (\lambda)$ is a variational action 
and $\lambda$ is a variational parameter to be
determined through the variational calculation. Then the partition function is given by
\begin{eqnarray}
Z &=& \int D[\bar{\psi} \psi] e^{-S[\bar{\psi}, \psi]} \nonumber \\
  &=& \int D[\bar{\psi} \psi] e^{-S_o [\bar{\psi}, \psi ; \lambda] -
  ( S[\bar{\psi}, \psi]- S_o [\bar{\psi}, \psi : \lambda] )} \nonumber \\
   &=& \left< e^{- \left( S\left[\zeta\frac{\partial}{\partial \eta},
   \frac{\partial}{\partial \bar{\eta}}\right]
   -S_o \left[\zeta\frac{\partial}{\partial \eta}, \frac{\partial}{\partial \bar{\eta}} ; \lambda \right] \right)}
    \right>_{S_o} \cdot \int D[\bar{\psi} \psi] e^{-S_o [\bar{\psi}, \psi : \lambda]}
      \nonumber \\
  &\geq& e^{- \left< \left( S\left[\zeta\frac{\partial}{\partial \eta},
   \frac{\partial}{\partial \bar{\eta}}\right] -S_o \left[\zeta\frac{\partial}{\partial \eta},
    \frac{\partial}{\partial \bar{\eta}} ; \lambda \right] \right) \right>_{S_o}} \cdot \int D[\bar{\psi} \psi]
     e^{-S_o [\bar{\psi}, \psi : \lambda]} \ .
\end{eqnarray}
The thermodynamic potential can be expressed likewise
\begin{equation}
\Omega \leq - \frac{1}{\beta} {\rm ln}\int D[\bar{\psi} \psi]
 e^{-S_o [\bar{\psi}, \psi : \lambda]} + \frac{1}{\beta}
\left<  S\left[\zeta\frac{\partial}{\partial \eta}, \frac{\partial}{\partial \bar{\eta}}\right]
 -S_o \left[\zeta\frac{\partial}{\partial \eta}, \frac{\partial}{\partial \bar{\eta}} ; \lambda \right]  \right>_{S_o} \ .
\end{equation}
We can obtain the variational result by maximizing the fourth line of Eq.(9) or minimizing 
the righthand side of Eq.(10) with $\lambda$.
Using the above expression, we now obtain the expressions for the variational perturbation
process for the partition function and the thermodynamic potential as follows,
\begin{eqnarray}
Z = & & \int D[\bar{\psi} \psi] e^{-S_o [\bar{\psi}, \psi : \lambda]}
 e^{- \left< \left( S\left[\zeta\frac{\partial}{\partial \eta}, \frac{\partial}{\partial \bar{\eta}}\right]
 -S_o \left[\zeta\frac{\partial}{\partial \eta}, \frac{\partial}{\partial \bar{\eta}} ; \lambda \right] \right)
\right>_{S_o}} \nonumber \\
  & & \cdot \left< e^{- \left( S\left[\zeta\frac{\partial}{\partial \eta},
   \frac{\partial}{\partial \bar{\eta}}\right] -S_o \left[\zeta\frac{\partial}{\partial \eta},
   \frac{\partial}{\partial \bar{\eta}} ; \lambda \right] \right)
         + \left<  S\left[\zeta\frac{\partial}{\partial \eta},
          \frac{\partial}{\partial \bar{\eta}}\right] -S_o \left[\zeta\frac{\partial}{\partial \eta},
          \frac{\partial}{\partial \bar{\eta}} ; \lambda \right]  \right>_{S_o}} \right>_{S_o} \ \ .
\end{eqnarray}
\begin{eqnarray}
\Omega = &-& \frac{1}{\beta} {\rm ln}\int D[\bar{\psi} \psi]
e^{-S_o [\bar{\psi}, \psi : \lambda]} + \frac{1}{\beta}
           \left<  S\left[\zeta\frac{\partial}{\partial \eta},
           \frac{\partial}{\partial \bar{\eta}}\right] -S_o \left[\zeta\frac{\partial}{\partial \eta},
            \frac{\partial}{\partial \bar{\eta}} ; \lambda \right]  \right>_{S_o} \nonumber \\
        &-&  \frac{1}{\beta} {\rm ln} \left< e^{- \left( S\left[\zeta\frac{\partial}{\partial \eta},
         \frac{\partial}{\partial \bar{\eta}}\right] -S_o \left[\zeta\frac{\partial}{\partial \eta},
         \frac{\partial}{\partial \bar{\eta}} ; \lambda \right] \right)
         + \left<  S\left[\zeta\frac{\partial}{\partial \eta},
         \frac{\partial}{\partial \bar{\eta}}\right] -S_o \left[\zeta\frac{\partial}{\partial \eta},
          \frac{\partial}{\partial \bar{\eta}} ; \lambda \right]  \right>_{S_o}} \right>_{S_o} \ \ .
\end{eqnarray}
The second lines of Eq.(11) and (12) are the 
 perturbations based on the variational
basis and, thus, provide systematic improvements over the variational result.

\section{Variational Perturbation on Interacting Fermion Systems and  Modified
         Wick's Theorem}

The above variational perturbation result on the thermodynamic potential can be applied
to any strongly correlated systems. In general, the variational result  
corresponds to the mean field value[15,16]. With the expressions given in Eq.(11) and (12), 
we can readily carry out systematic
improvements over the mean field result,
 since the renormalized perturbation term is 
generally
much smaller than the original one. To see this point clearly, we apply the principle
to an interacting fermion gas system through two-body interactions. 
In the process, we derive a modified
Wick's theorem, which removes the problem of double counting of higher order terms.
We also obtain explicit expressions for the thermodynamic potential.

The partition function for two-body interacting fermion systems
is generally given by
\begin{equation}
Z = \int D[\bar{\psi}\psi] \ e^{-G_{ij}^{o^{-1}}
 \bar{\psi}_{i} \psi_{j} - V_{ijkl}
\bar{\psi}_{i} \bar{\psi}_{j} \psi_{k} \psi_{l} } \ .
\end{equation}
where the subscript stands for all possible quantum numbers and an imaginary time $\tau$ ;
$i = (k \sigma \tau)$.
The  integration or summation symbols are omitted by the summation convention.
$G^o $ is a bare Green's function matrix and $V$ is an interaction tensor. For example,
for free fermions, $G^o$ has a form;
\begin{eqnarray}
G_{ij}^{o^{-1}} \rightarrow \langle k \sigma \tau | G^{o^{-1}}| k' \sigma' \tau' \rangle
=(\partial_{\tau} + \hbar^2 k^2 /2m - \mu) \delta (\tau-\tau') \delta_{kk'} 
\delta_{\sigma \sigma'} \ .
\end{eqnarray}
The partition function can be 
rewritten using the variational Green's function $G$ ;
\begin{eqnarray}
Z &=& \int D[\bar{\psi}\psi] \ e^{-G_{ij}^{-1}
\bar{\psi}_{i} \psi_{j}-( G_{ij}^{o^{-1}} - G_{ij}^{-1} )
 \bar{\psi}_{i} \psi_{j} -V_{ijkl}
\bar{\psi}_{i} \bar{\psi}_{j} \psi_{k} \psi_{l} - \psi_{i} \bar{\eta}_{i}
- \eta_{i} \bar{\psi}_{i}} |_{\bar{\eta}, \eta =0} \nonumber \\
&=& e^{{\rm ln  Det} G^{-1} + ( G_{ij}^{o^{-1}} - G_{ij}^{-1} ) \frac{\partial}{\partial \eta_{i}}
\frac{\partial}{\partial \bar{\eta}_{j}} -V_{ijkl} \frac{\partial}{\partial \eta_{i}}
\frac{\partial}{\partial \eta_{j}} \frac{\partial}{\partial \bar{\eta}_{k}}
\frac{\partial}{\partial \bar{\eta}_{l} }} 
 e^{\bar{\eta}_{i} G_{ij} \eta_{j}} |_{\bar{\eta}, \eta =0} \ \ .
\end{eqnarray}
Here we used the Gaussian integral for the Grassmann variables [3];
\begin{eqnarray}
\int D[\bar{\psi}\psi]  e^{- \bar{\psi}_{i}
G_{ij}^{-1} \psi_{j} + \bar{\eta}_{i} \psi_{i} +
\bar{\psi}_{i} \eta_{i}}  = {\rm Det} G^{-1} e^{ \bar{\eta}_{i}
G_{ij} \eta_{j}} \ .
\end{eqnarray}
Using the shorthand notation of Eq.(7), the partition function is expressed
\begin{equation}
Z = e^{{\rm ln  Det} G^{-1}} \left< e^{ ( G_{ij}^{o^{-1}} - G_{ij}^{-1} ) \frac{\partial}{\partial \eta_{i}}
\frac{\partial}{\partial \bar{\eta}_{j}} -V_{ijkl} \frac{\partial}{\partial \eta_{i}}
\frac{\partial}{\partial \eta_{j}} \frac{\partial}{\partial \bar{\eta}_{k}}
\frac{\partial}{\partial \bar{\eta}_{l} } } \right>_{G} \ \ .
\end{equation}
Using the variational principle  discussed above, 
 we obtain for the partition function
\begin{equation}
Z \geq \bar{Z} = e^{{\rm ln  Det} G^{-1} + ( G_{ij}^{o^{-1}} - G_{ij}^{-1} )
\left< \frac{\partial}{\partial \eta_{i}}
\frac{\partial}{\partial \bar{\eta}_{j}} \right>_{G} -V_{ijkl}
\left< \frac{\partial}{\partial \eta_{i}}
\frac{\partial}{\partial \eta_{j}} \frac{\partial}{\partial \bar{\eta}_{k}}
\frac{\partial}{\partial \bar{\eta}_{l} }  \right>_{G} } \ \ .
\end{equation}
It is now necessary to obtain the Green's function, $G$, which appears in the 
variational process. For this purpose, we consider the variational thermodynamic
potential.
\begin{eqnarray}
\Omega \leq \bar{\Omega} =&-& \frac{1}{\beta}{\rm ln
Det}G^{-1}-\frac{1}{\beta} ( G_{ij}^{o^{-1}} - G_{ij}^{-1} )
\left< \frac{\partial}{\partial \eta_{i}}
\frac{\partial}{\partial \bar{\eta}_{j}} \right>_{G} \nonumber \\
 &+& \frac{1}{\beta}
V_{ijkl}
\left< \frac{\partial}{\partial \eta_{i}}
\frac{\partial}{\partial \eta_{j}} \frac{\partial}{\partial \bar{\eta}_{k}}
\frac{\partial}{\partial \bar{\eta}_{l} }  \right>_{G} \nonumber \\
= &-&\frac{1}{\beta}{\rm ln
Det}G^{-1}-\frac{1}{\beta} ( G_{ij}^{o^{-1}} - G_{ij}^{-1} )
G_{ji} \nonumber \\
&+&\frac{1}{\beta}V_{ijkl}( G_{kj} G_{li} - G_{ki} G_{lj} ) \ \ .
\end{eqnarray}
Minimizing $\bar{\Omega}$
against the variational Green's function $G$, we obtain
\begin{equation}
G_{ba}^{-1}- G_{b a}^{o^{-1}} + 2(V_{ibaj} - V_{biaj}) G_{ji}
= 0 \ .
\end{equation}
In deriving the above result, the symmetry property of $V_{ijkl} = V_{jilk}$
is used. Defining $\left< b| \Sigma |a \right> \equiv 2 ( V_{ibaj} - V_{biaj}) G_{ji}$ ,
we obtain
\begin{equation}
G^{-1} - G^{o^{-1}} + \Sigma = 0 \ .
\end{equation}
Here, we note that  $G$ corresponds to the
self-consistent Hartree-Fock Green's function for the interacting fermion system.
Using this variational Green's function, we rewrite the partition function
as follows,
\begin{eqnarray}
&&Z = e^{{\rm ln  Det} G^{-1} + ( G_{ij}^{o^{-1}} - G_{ij}^{-1} )
\left< \frac{\partial}{\partial \eta_{i}}
\frac{\partial}{\partial \bar{\eta}_{j}} \right>_{G} -V_{ijkl}
\left< \frac{\partial}{\partial \eta_{i}}
\frac{\partial}{\partial \eta_{j}} \frac{\partial}{\partial \bar{\eta}_{k}}
\frac{\partial}{\partial \bar{\eta}_{l} }  \right>_{G} } \nonumber \\
&&\cdot
 \left< e^{ ( G_{ij}^{o^{-1}} - G_{ij}^{-1} )
 \left( \frac{\partial}{\partial \eta_{i}}
\frac{\partial}{\partial \bar{\eta}_{j}}- \left<
\frac{\partial}{\partial \eta_{i}}
\frac{\partial}{\partial \bar{\eta}_{j}} \right>_{G} \right)
 -V_{ijkl} \left( \frac{\partial}{\partial \eta_{i}}
\frac{\partial}{\partial \eta_{j}} \frac{\partial}{\partial \bar{\eta}_{k}}
\frac{\partial}{\partial \bar{\eta}_{l} } - \left<
\frac{\partial}{\partial \eta_{i}}
\frac{\partial}{\partial \eta_{j}} \frac{\partial}{\partial \bar{\eta}_{k}}
\frac{\partial}{\partial \bar{\eta}_{l} } \right>_{G} \right)} \right>_{G} \ \ .
\end{eqnarray}
The second line in the above equation is the renormalized perturbation
contribution and can be simplified as follows,
\begin{eqnarray}
&& ( G_{ij}^{o^{-1}} - G_{ij}^{-1} )
 \left( \frac{\partial}{\partial \eta_{i}}
\frac{\partial}{\partial \bar{\eta}_{j}}- \left<
\frac{\partial}{\partial \eta_{i}}
\frac{\partial}{\partial \bar{\eta}_{j}} \right>_{G} \right) \nonumber \\
& & \ \ \ -V_{ijkl} \left( \frac{\partial}{\partial \eta_{i}}
\frac{\partial}{\partial \eta_{j}} \frac{\partial}{\partial \bar{\eta}_{k}}
\frac{\partial}{\partial \bar{\eta}_{l} } - \left<
\frac{\partial}{\partial \eta_{i}}
\frac{\partial}{\partial \eta_{j}} \frac{\partial}{\partial \bar{\eta}_{k}}
\frac{\partial}{\partial \bar{\eta}_{l} } \right>_{G} \right) \nonumber \\
& &= ( V_{kijl} + V_{iklj} - V_{ikjl} - V_{kilj} )
 \left(  \frac{\partial}{\partial \eta_{i}}
\frac{\partial}{\partial \bar{\eta}_{j}} -G_{ji} \right) G_{lk} \nonumber \\
& & \ \ \ - V_{ijkl} \left(  \frac{\partial}{\partial \eta_{i}}
\frac{\partial}{\partial \eta_{j}} \frac{\partial}{\partial \bar{\eta}_{k}}
\frac{\partial}{\partial \bar{\eta}_{l} }
- G_{kj} G_{li} + G_{ki} G_{lj} \right) \nonumber \\
& &= -V_{ijkl} \left( \frac{\partial}{\partial \eta_{i}}
\frac{\partial}{\partial \eta_{j}} \frac{\partial}{\partial \bar{\eta}_{k}}
\frac{\partial}{\partial \bar{\eta}_{l} } - G_{li} \frac{\partial}{\partial \eta_{j}}
\frac{\partial}{\partial \bar{\eta}_{k}}
-G_{kj} \frac{\partial}{\partial \eta_{i}} \frac{\partial}{\partial \bar{\eta}_{l}} \right.
   \nonumber \\
& &  \ \ \ \left. +G_{lj} \frac{\partial}{\partial \eta_{i}} \frac{\partial}{\partial \bar{\eta}_{k}}
+G_{ki} \frac{\partial}{\partial \eta_{j}}\frac{\partial}{\partial \bar{\eta}_{l}}
+G_{li} G_{kj} - G_{lj} G_{ki} \right) \nonumber \\
& &\equiv -V_{ijkl} \left( \frac{\partial}{\partial \eta_{i}}
\frac{\partial}{\partial \eta_{j}} \frac{\partial}{\partial \bar{\eta}_{k}}
\frac{\partial}{\partial \bar{\eta}_{l} }  \right)'
\end{eqnarray}
In deriving the above relation, we used the symmetry of the potential $V_{ijkl}$ and
Eq.(20). 
Here the primed operation $( \cdot \cdot \cdot )'$ is defined to be
\begin{eqnarray}
\left( \frac{\partial}{\partial \eta_{i}}
\frac{\partial}{\partial \eta_{j}} \frac{\partial}{\partial \bar{\eta}_{k}}
 \frac{\partial}{\partial \bar{\eta}_{l} }  \right)' 
&\equiv&  \left( \frac{\partial}{\partial \eta_{i}}
\frac{\partial}{\partial \eta_{j}} \frac{\partial}{\partial \bar{\eta}_{k}}
\frac{\partial}{\partial \bar{\eta}_{l} } - G_{li} \frac{\partial}{\partial \eta_{j}}
\frac{\partial}{\partial \bar{\eta}_{k}}
-G_{kj} \frac{\partial}{\partial \eta_{i}} \frac{\partial}{\partial \bar{\eta}_{l}} \right.
   \nonumber \\
& &  \ \ \ \left. +G_{lj} \frac{\partial}{\partial \eta_{i}} \frac{\partial}{\partial \bar{\eta}_{k}}
+G_{ki} \frac{\partial}{\partial \eta_{j}}\frac{\partial}{\partial \bar{\eta}_{l}}
+G_{li} G_{kj} - G_{lj} G_{ki} \right) \ .
\end{eqnarray}
Using this primed operation and Eq. (21), the partition function is greatly simplified.
\begin{eqnarray}
Z = e^{- {\rm tr ln}G + \frac{1}{2} {\rm tr} \Sigma G}
   \left< e^{-V_{ijkl} \left( \frac{\partial}{\partial \eta_{i}}
\frac{\partial}{\partial \eta_{j}} \frac{\partial}{\partial \bar{\eta}_{k}}
\frac{\partial}{\partial \bar{\eta}_{l} }   \right)'} \right>_{G}
\end{eqnarray}
In order to carry out the perturbation, it is necessary to evaluate terms like
\begin{eqnarray}
&& \left<   \left( \frac{\partial}{\partial \eta_{i_1 }}
\frac{\partial}{\partial \eta_{j_1 }} \frac{\partial}{\partial \bar{\eta}_{k_1 }}
\frac{\partial}{\partial \bar{\eta}_{l_1 } } \right)' \cdot \cdot \cdot
 \left( \frac{\partial}{\partial \eta_{i_n }}
\frac{\partial}{\partial \eta_{j_n }} \frac{\partial}{\partial \bar{\eta}_{k_n }}
\frac{\partial}{\partial \bar{\eta}_{l_n } }  \right)'   \right>_{G}
\nonumber \\
&&= \left( \frac{\partial}{\partial \eta_{i_1 }}
\frac{\partial}{\partial \eta_{j_1 }} \frac{\partial}{\partial \bar{\eta}_{k_1 }}
\frac{\partial}{\partial \bar{\eta}_{l_1 } } \right)' \cdot \cdot \cdot
 \left( \frac{\partial}{\partial \eta_{i_n }}
\frac{\partial}{\partial \eta_{j_n }} \frac{\partial}{\partial \bar{\eta}_{k_n }}
\frac{\partial}{\partial \bar{\eta}_{l_n } }  \right)'
 e^{\bar{\eta}_{i} G_{ij} \eta_j } |_{\bar{\eta}=\eta=0} \ .
\end{eqnarray}
With the prime defined in Eq.(24), we cannot apply Wick's theorem 
directly to the above expansion. Therefore, we first study contraction rules
under the primed operation. We rewrite Eq.(24) as follows,
\begin{eqnarray}
\begin{picture}(300,60)
\put(162,51){\line(0,-1){5}}
\put(162,51){\line(1,0){26}}
\put(188,51){\line(0,-1){5}}

\put(233,51){\line(0,-1){5}}
\put(233,51){\line(1,0){10}}
\put(243,51){\line(0,-1){5}}

\put(296,51){\line(0,-1){5}}
\put(296,51){\line(1,0){18}}
\put(314,51){\line(0,-1){5}}

\put(113,16){\line(0,-1){5}}
\put(113,16){\line(1,0){18}}
\put(131,16){\line(0,-1){5}}

\put(177,20){\line(0,-1){9}}
\put(177,20){\line(1,0){28}}
\put(205,20){\line(0,-1){9}}

\put(186,16){\line(0,-1){5}}
\put(186,16){\line(1,0){10}}
\put(196,16){\line(0,-1){5}}

\put(240,16){\line(0,-1){5}}
\put(240,16){\line(1,0){18}}
\put(258,16){\line(0,-1){5}}

\put(248,20){\line(0,-1){9}}
\put(248,20){\line(1,0){18}}
\put(266,20){\line(0,-1){9}}

\put(-8,35){ $
 \left( \frac{\partial}{\partial \eta_i } \frac{\partial}{\partial \eta_j }
\frac{\partial}{\partial \bar{\eta}_k } \frac{\partial}{\partial \bar{\eta}_l }
 \right)' =
 ( \ i \ j \ \bar{k} \ \bar{l} \ ) - ( \ i \ j \ \bar{k} \ \bar{l} \ ) - ( \ i \  j \ \bar{k} \ \bar{l} \ )
  - ( \ i \ j \ \bar{k} \ \bar{l} \ ) $}
\put(90,0){ $ - ( \ i \ j \ \bar{k} \ \bar{l} \ ) + ( \ i \ j \ \bar{k} \ \bar{l} \ ) +
 ( \ i \ j \ \bar{k} \ \bar{l} \ ) \  , $}
\end{picture}
\end{eqnarray}
where we used shorthand notations; $\frac{\partial}{\partial \eta_i } \rightarrow i $ and 
$\frac{\partial}{\partial \bar{\eta}_{i} } \rightarrow \bar{i}$, and 
the connected pairs represent the contractions;
\begin{picture}(50,16)
\put(5,16){\line(0,-1){5}}
\put(5,16){\line(1,0){10}}
\put(15,16){\line(0,-1){5}}
\put(0,0){ $ i \ \bar{j} \equiv G_{ji} $ }
\end{picture}.
We call one $( \cdot \cdot \cdot )'$ a unit cell and mark a contraction 
with other unit cell elements with dots and a intracell contraction with lines.
With these notations, terms in Eq.(27) appear
\begin{eqnarray} 
&&
\begin{picture}(300,70)
\put(151,51){\line(0,-1){5}}
\put(151,51){\line(1,0){28}}
\put(179,51){\line(0,-1){5}}

\put(224,51){\line(0,-1){5}}
\put(224,51){\line(1,0){10}}
\put(234,51){\line(0,-1){5}}

\put(287,51){\line(0,-1){5}}
\put(287,51){\line(1,0){18}}
\put(305,51){\line(0,-1){5}}

\put(93,16){\line(0,-1){5}}
\put(93,16){\line(1,0){18}}
\put(111,16){\line(0,-1){5}}

\put(155,20){\line(0,-1){9}}
\put(155,20){\line(1,0){28}}
\put(182,20){\line(0,-1){9}}

\put(164,16){\line(0,-1){5}}
\put(164,16){\line(1,0){10}}
\put(174,16){\line(0,-1){5}}

\put(219,16){\line(0,-1){5}}
\put(219,16){\line(1,0){18}}
\put(237,16){\line(0,-1){5}}

\put(228,20){\line(0,-1){9}}
\put(228,20){\line(1,0){18}}
\put(246,20){\line(0,-1){9}}

\put(89,48){\circle*{2}}
\put(98,48){\circle*{2}}
\put(107,48){\circle*{2}}
\put(116,48){\circle*{2}}

\put(160,48){\circle*{2}}
\put(169,48){\circle*{2}}

\put(215,48){\circle*{2}}
\put(242,48){\circle*{2}}

\put(278,48){\circle*{2}}
\put(296,48){\circle*{2}}

\put(102,13){\circle*{2}}
\put(120,13){\circle*{2}}

\put(10,35){ $ ( \ i \ j \ \bar{k} \ \bar{l} \ ) =
 ( \ i \ j \ \bar{k} \ \bar{l} \ ) + ( \ i \ j \ \bar{k} \ \bar{l} \ ) + ( \ i \  j \ \bar{k} \ \bar{l} \ )
  + ( \ i \ j \ \bar{k} \ \bar{l} \ ) $}
\put(70,0){ $ + ( \ i \ j \ \bar{k} \ \bar{l} \ ) + ( \ i \ j \ \bar{k} \ \bar{l} \ ) +
 ( \ i \ j \ \bar{k} \ \bar{l} \ ) , $}
\end{picture}
\nonumber \\
&&
\begin{picture}(100,40)
\put(23,17){\line(0,-1){5}}
\put(23,17){\line(1,0){28}}
\put(52,17){\line(0,-1){5}}

\put(96,17){\line(0,-1){5}}
\put(96,17){\line(1,0){28}}
\put(124,17){\line(0,-1){5}}

\put(161,20){\line(0,-1){9}}
\put(161,20){\line(1,0){28}}
\put(189,20){\line(0,-1){9}}

\put(170,16){\line(0,-1){5}}
\put(170,16){\line(1,0){10}}
\put(180,16){\line(0,-1){5}}

\put(106,13){\circle*{2}}
\put(115,13){\circle*{2}}

\put(0,0){ $ - ( \ i \ j \ \bar{k} \ \bar{l} \ ) =
 - ( \ i \ j \ \bar{k} \ \bar{l} \ ) - ( \ i \ j \ \bar{k} \ \bar{l} \ ) , $ }
\end{picture}
\nonumber \\
&&
\begin{picture}(100,40)
\put(33,17){\line(0,-1){5}}
\put(33,17){\line(1,0){10}}
\put(43,17){\line(0,-1){5}}

\put(106,17){\line(0,-1){5}}
\put(106,17){\line(1,0){10}}
\put(116,17){\line(0,-1){5}}

\put(162,20){\line(0,-1){9}}
\put(162,20){\line(1,0){28}}
\put(190,20){\line(0,-1){9}}

\put(171,16){\line(0,-1){5}}
\put(171,16){\line(1,0){10}}
\put(181,16){\line(0,-1){5}}

\put(97,13){\circle*{2}}
\put(124,13){\circle*{2}}

\put(0,0){ $ - ( \ i \ j \ \bar{k} \ \bar{l} \ ) =
 - ( \ i \ j \ \bar{k} \ \bar{l} \ ) - ( \ i \ j \ \bar{k} \ \bar{l} \ ) , $ }
\end{picture}
\nonumber \\
&&
\begin{picture}(100,40)
\put(33,17){\line(0,-1){5}}
\put(33,17){\line(1,0){18}}
\put(52,17){\line(0,-1){5}}

\put(106,17){\line(0,-1){5}}
\put(106,17){\line(1,0){18}}
\put(124,17){\line(0,-1){5}}

\put(170,20){\line(0,-1){9}}
\put(170,20){\line(1,0){18}}
\put(188,20){\line(0,-1){9}}

\put(161,16){\line(0,-1){5}}
\put(161,16){\line(1,0){18}}
\put(179,16){\line(0,-1){5}}

\put(97,13){\circle*{2}}
\put(115,13){\circle*{2}}

\put(0,0){ $ - ( \ i \ j \ \bar{k} \ \bar{l} \ ) =
 - ( \ i \ j \ \bar{k} \ \bar{l} \ ) - ( \ i \ j \ \bar{k} \ \bar{l} \ ) , $ }
\end{picture}
\nonumber \\
&&
\begin{picture}(100,40)
\put(23,17){\line(0,-1){5}}
\put(23,17){\line(1,0){18}}
\put(42,17){\line(0,-1){5}}

\put(97,17){\line(0,-1){5}}
\put(97,17){\line(1,0){18}}
\put(115,17){\line(0,-1){5}}

\put(170,20){\line(0,-1){9}}
\put(170,20){\line(1,0){18}}
\put(188,20){\line(0,-1){9}}

\put(161,16){\line(0,-1){5}}
\put(161,16){\line(1,0){18}}
\put(179,16){\line(0,-1){5}}

\put(106,13){\circle*{2}}
\put(124,13){\circle*{2}}

\put(0,0){ $ - ( \ i \ j \ \bar{k} \ \bar{l} \ ) =
 - ( \ i \ j \ \bar{k} \ \bar{l} \ ) - ( \ i \ j \ \bar{k} \ \bar{l} \ ) . $ }
\end{picture}
\end{eqnarray}
Now adding all contributions within one cell, we find that all intracell contractions
cancel and disappear, thus making the perturbation with the variational basis
simple and straightforward. 
The final expression for each primed operation in Eq.(26) appears 
\begin{equation}
\begin{picture}(100,35)

\put(46,13){\circle*{2}}
\put(55,13){\circle*{2}}
\put(64,13){\circle*{2}}
\put(73,13){\circle*{2}}

\put(-60,0){ $ \left( \frac{\partial}{\partial \eta_i } \frac{\partial}{\partial \eta_j }
\frac{\partial}{\partial \bar{\eta}_k } \frac{\partial}{\partial \bar{\eta}_l }
 \right)' =
 ( \ i \ j \ \bar{k} \ \bar{l} \ ) \ . $  }
\end{picture}
\end{equation}
Therefore, the contraction rule for higher terms now becomes
\begin{eqnarray}
\begin{picture}(100,100)

\put(-36,48){\circle*{2}}
\put(-23,48){\circle*{2}}
\put(-10,48){\circle*{2}}
\put(3,48){\circle*{2}}

\put(50,48){\circle*{2}}
\put(64,48){\circle*{2}}
\put(78,48){\circle*{2}}
\put(92,48){\circle*{2}}

\put(-70,70){ $ \left<   \left( \frac{\partial}{\partial \eta_{i_1 }}
\frac{\partial}{\partial \eta_{j_1 }} \frac{\partial}{\partial \bar{\eta}_{k_1 }}
\frac{\partial}{\partial \bar{\eta}_{l_1 } } \right)' \cdot \cdot \cdot
 \left( \frac{\partial}{\partial \eta_{i_n }}
\frac{\partial}{\partial \eta_{j_n }} \frac{\partial}{\partial \bar{\eta}_{k_n }}
\frac{\partial}{\partial \bar{\eta}_{l_n } }  \right)'   \right>_{G} $ }
\put(-70,35){ $ = \left<   ( \ i_1  \ j_1 \ \bar{k}_1  \ \bar{l}_1  \ ) \cdot \cdot \cdot
 ( \ i_n \ j_n \ \bar{k}_n  \ \bar{l}_n  \ )  \right>_{G} $ }
\put(-70,5){  $=$ Sum of all contractions between different cells.  }
\end{picture}
\end{eqnarray}
This modification of Wick's theorem represents nothing but a requirement that the
selfenergy contributions which are included in the variational Green's function
should not contribute again in the perturbation calculation. 
This new rule allows one to carry out higher order calculations without worrying
about the double counting problem. The perturbation calculations are easily tractable
by the diagramatic method. 
The unit cell with an interaction, 
$V_{ijkl}\left( \frac{\partial}{\partial \eta_{i}}
\frac{\partial}{\partial \eta_{j}} \frac{\partial}{\partial \bar{\eta}_{k}}
\frac{\partial}{\partial \bar{\eta}_{l} }   \right)' $ is shown in Fig.1 (a). 
The modified Wick's theorem stipulates that diagrams with intracell contractions do not
contribute to the thermodynamic potential. Examples of such diagrams are shown in Fig.1 (b).

Since the linked cluster theorem is equally applicable in the present formulation,
the thermodynamic potential can now readily be written
\begin{eqnarray}
\Omega =\frac{1}{\beta} {\rm tr ln}G -\frac{1}{2 \beta} {\rm tr} \Sigma G
-\frac{1}{\beta} \left< e^{-V_{ijkl} \left( \frac{\partial}{\partial \eta_{i}}
\frac{\partial}{\partial \eta_{j}} \frac{\partial}{\partial \bar{\eta}_{k}}
\frac{\partial}{\partial \bar{\eta}_{l} }   \right)'} \right>_{G, \ C}
\end{eqnarray}
where $C$ indicates the connected diagrams only.
The first two terms are the variational result and the last is 
the renormalized perturbation.
 For example, 
 the second order perturbation terms are given as follows;
\begin{eqnarray}
&&-\frac{1}{\beta} \cdot \frac{1}{2} V_{ijkl} V_{i'j'k'l'} \left
\langle  \left( \frac{\partial}{\partial \eta_{i}}
            \frac{\partial}{\partial \eta_{j}} \frac{\partial}{\partial \bar{\eta}_{k}}
            \frac{\partial}{\partial \bar{\eta}_{l} }   \right)'
 \left( \frac{\partial}{\partial \eta_{i'}}
            \frac{\partial}{\partial \eta_{j'}} \frac{\partial}{\partial \bar{\eta}_{k'}}
            \frac{\partial}{\partial \bar{\eta}_{l'} }   \right)'  \right \rangle_{G, \
            C} \nonumber \\
&&= -\frac{1}{\beta}  V_{ijkl} V_{i'j'k'l'} G_{l'i}G_{k'j}G_{kj'}G_{li'}
\nonumber \\
&& \ \ \ +\frac{1}{\beta}  V_{ijkl} V_{i'j'k'l'} G_{l'i}G_{k'j}G_{ki'}G_{lj'} \ ,
\end{eqnarray}
which correspond to the diagrams in Fig.2 (a).

\section{a model calculation}

A sample calculation was carried out to test the efficiency of the present method
using a three dimensional nucleon gas system interacting through the 
two-body Yukawa-type potential.
The model Hamiltonian is given by
\begin{equation}
H = \sum_{k \sigma} \epsilon_k  a_{k \sigma}^{\dagger} a_{k \sigma} + \frac{1}{2}
    \sum_{k k' q \sigma \sigma'} \hspace{-0.25cm} ^{'}  v(q) a_{k \sigma}^{\dagger} a_{k' \sigma'}^{\dagger}
    a_{k'-q \sigma'} a_{k+q \sigma} \ ,
\end{equation}
where
\begin{eqnarray}
 \epsilon_{k} &=& \frac{\hbar^2 k^2}{2 m_n }  \nonumber \\
 v(q) &=& \frac{4 \pi \gamma }{V ( q^2 + \lambda^2 )} \ .
\end{eqnarray}
Here, $m_n$ is the nucleon mass and $\gamma$ and $\lambda$ are model parameters 
which should be fixed for
numerical calculations. The primed summation indicates that the $q=0$ term is 
renormalized away because
this  divergent energy term 
has the same contribution for both the conventional and the variational perturbation methods. 
With this model Hamiltonian, we calculate the ground state
energy at zero temperature using both the variational perturbation theory
and the conventional perturbation theory up to the second order. 
The bare Green's function and the interaction of this model is expressed as
\begin{eqnarray}
 & & G_{ij}^{o^{-1}} \rightarrow  G_{ k \sigma \tau ,  k' \sigma' \tau'}^{o^{-1}}
 = ( \partial_\tau + \epsilon_{k} - \mu ) \delta(\tau - \tau') \delta_{k k'}
 \delta_{\sigma \sigma'}  \ , \\
 & & V_{ijkl} \rightarrow V_{ k_{1} \sigma_{1} \tau_{1}, k_{2} \sigma_{2} \tau_{2},
 k_{3} \sigma_{3} \tau_{3}, k_{4} \sigma_{4} \tau_{4}} \nonumber \\
& & \ \ \ \ \ \ \ \ \ \
= \frac{1}{2} v( | k_4 - k_1 | ) \delta_{k_1 + k_2 , k_3 + k_4} \delta_{\sigma_1
\sigma_4}
\delta_{\sigma_2 \sigma_3} 
 \delta(\tau_1 - \tau_2) \delta(\tau_3 - \tau_4)
\delta(\tau_2 - \tau_3) \ ,
\end{eqnarray}
where $\mu$ is the chemical potential.
 Denoting $G_{ k \sigma \tau ,  k \sigma \tau'}$ as
$G_{k \sigma}(\tau -\tau')$ for a homogeneous system,
the equation of $G$, Eq.(21) is expressed as follows,
\begin{eqnarray}
G_{k \sigma}(\tau -\tau') 
= G_{k \sigma}^{o}(\tau -\tau') 
 - \int d\tau'' G_{k \sigma}^{o} (\tau -\tau'')
 \sum_{q} v(q) G_{k+q \sigma}(0)  G_{k \sigma}(\tau''-\tau') \ .
\end{eqnarray} 
Fourier transforming of this equation with respect to time, we easily obtain
the variational
Green's function $G_{k \sigma}(i \omega_n )$  as
\begin{eqnarray}
G_{k \sigma}(i \omega_n ) =\frac{-1}{i \omega_n - 
( \tilde{\epsilon}_{k \sigma}  - \mu )} \ ,
\end{eqnarray}
where $\omega_n$ is the fermion Matsubara frequency, $\frac{(2n+1)\pi}{\beta}$ and 
$\tilde{\epsilon}_{k \sigma }$ is the renormalized nucleon energy value given by
$\epsilon_k  - \sum_{q} v(q) n_{k+q \sigma}$.   
$n_{k \sigma}$ is the Fermi distribution function,  
 $\frac{1}{e^{\beta(\tilde{\epsilon}_{k \sigma}-\mu)}+1}$
 with the renormalized energy. 
Therefore, the variational result of the thermodynamic potential, $\bar{\Omega}_{min}$ 
is
\begin{eqnarray}
 \bar{\Omega}_{min} &=& \frac{1}{\beta} {\rm tr ln}G -\frac{1}{2 \beta}
                {\rm tr} \Sigma G  \nonumber \\
 &=& - \frac{1}{\beta} \sum_{k \sigma} {\rm ln} \left( 1 + e^{-\beta (\epsilon_{k}
  - \sum_{q} v(q) n_{k+q \sigma} - \mu )} \right)
  + \frac{1}{2} \sum_{k q \sigma} v(q) n_{k+q \sigma} n_{k \sigma} \ .
\end{eqnarray}
The first order contribution is absent due to the modified Wick's theorem and the 
relevant second order diagrams are shown in Fig.2 (a), while conventional diagrams
of the same order are also shown in Fig.2 (b) for the comparison, where 
 double lines represent the variational $G$ and single lines the bare $G^o$.
The details of the calculations are exactly the same for both cases except that the
double line Green's function has the one particle energy, $\tilde{\epsilon}_{k \sigma }$
instead of the bare energy, ${\epsilon}_{k }$.
For the nucleon gas interacting with the 
potential given in Eq.(34), at zero temperature, the energy $\tilde{\epsilon}_{k \sigma}$
can be evaluated analytically to be given by
\begin{eqnarray}
\tilde{\epsilon}_{k \sigma} 
= \frac{\gamma}{2 r} \left[ \left( \frac{9 \pi}{8} \right)^{\frac{2}{3}}
\frac{\hbar^2}{\gamma r m_n } \left( \frac{k}{k_{F}^0 } \right)^2
 -\frac{4}{\pi}  \left( \frac{9 \pi}{8} \right)^{\frac{1}{3}} 
F\left( \frac{k}{ k_{F}^0 } , \frac{\lambda r }{ ( {9 \pi}/{8} )^{{1}/{3}} }\right)  \right] \ ,
\end{eqnarray}
where $r$ is the average distance between nucleons which is 
defined by $V= \frac{4}{3} \pi r^3 N$ with the volume $V$ and the 
number of nucleons $N$ of the system. $k_{F}^0$ is the Fermi wave vector and
$F(x, b)$ is given by
\begin{eqnarray}
F(x, b) = \frac{1}{2} + \frac{1+b^2 -x^2}{8 x} {\rm ln}
\frac{b^2 +(x+1)^2 }{b^2 +(x-1)^2} - \frac{1}{2} b \left[ \tan^{-1}\frac{1-x}{b}+
 \tan^{-1}\frac{1+x}{b} \right] \ .
\end{eqnarray}
Using the above prescriptions and following standard text book steps [1],
we calculated the total energy at zero temperature numerically. The result
is shown in Fig.3 and compared to the conventional calculation. Since the 
exact solution is not available, it is not possible to access the accuracy of
the present method directly. It is known that, for the homogeneous fermion gas,
the variational Hartree-Fock result corresponds to the standard first order calculation [18].
Therefore, a convergence test can be carried out by comparing the second order
contributions from both methods with the  Hartree-Fock result.
The result in Fig.3 clearly shows that the magnitude of the second order contribution
is substantially reduced in the variational perturbation calculation.
For example, at $r = 2.5$ Fermi, the renormalized contribution is 60.2\% of the bare contribution.
This renormalization clearly originates from the higher order terms included in the
variational perturbation process [12].

The simple Yukawa-type potential considered here allows an analytic expression
for the variational one-particle energy $\tilde{\epsilon}_{k \sigma}$. 
However, in real systems, the interaction between nucleons is more complicated
and, thus, generally does not yield a closed expression. We note that, although
an analytic expression for $\tilde{\epsilon}_{k \sigma}$ is convenient to carry out
the calculation, a numerical form of $\tilde{\epsilon}_{k \sigma}$ does not pose
a fundamental difficulty, once whole spectrum of the energy values is known.

\section{Summary}

In conclusion, we have presented a variational perturbation scheme for many particle
systems using the functional integral formulation. It is shown that the Green's
function obtained through the variational process corresponds to the Hartree-Fock
Green's function. A modified Wick's theorem for renormalized perturbation based on
the variational Green's function is obtained.
A model calculation with a nucleon gas interacting with the Yukawa type potential
shows that the present formalism provides an efficient and convenient formalism
to study strongly correlated systems.

\acknowledgements
This work was partly supported by the Korea Research Foundation(99-005-D00011)
and by the Brain Korea 21 Project.

\begin{figure}
\caption{(a) Unit cell to construct Feynmann diagrams,
         (b) Diagrams which do not appear in the variational perturbation calculation.
             The double line indicates the variational Green's
             function.} 
\end{figure}

\begin{figure}
\caption{Second order diagrams (a) for the present variational perturbation scheme and
         (b) for the conventional perturbation theory. Note that the last diagram in
         (b) does not appear in (a) due to the modified Wick's theorem.} 
\end{figure}

\begin{figure}
\caption{The ground state energy per particle calculated up to the second order; (a) solid line
   represents the present variational perturbation result, (b) dotted line
   the conventional perturbation upto the second order,  and (c) dashed line the
   Hartree-Fock result. Here, $\frac{\hbar}{m_{\pi} c}$ is the Compton wavelength of the pion [17].} 
\end{figure}

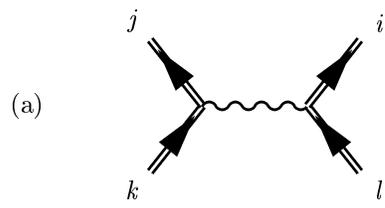
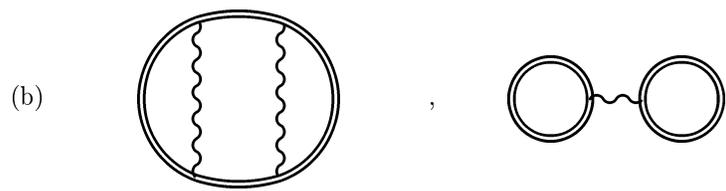
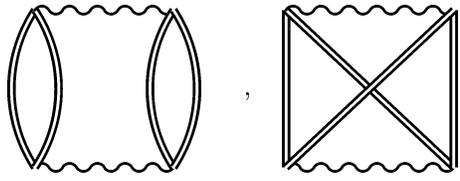
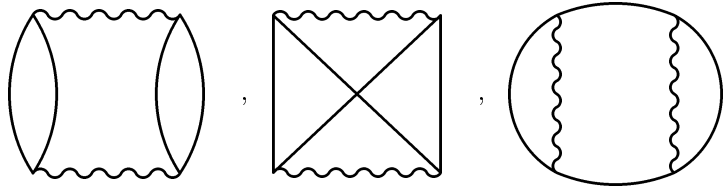
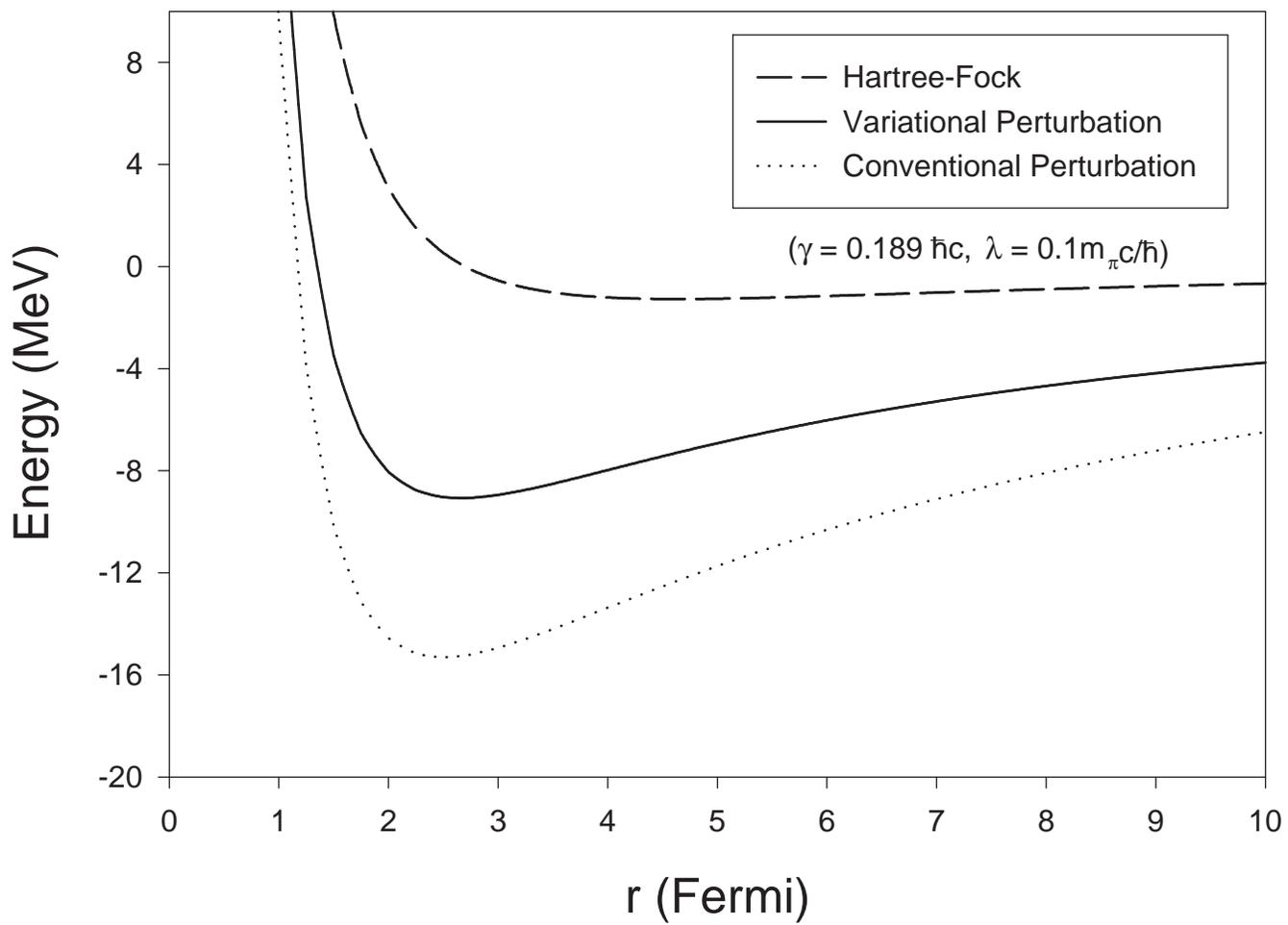
\end{document}